\documentclass[aip,reprint,jcp]{revtex4-1}

\usepackage{graphicx}
\usepackage{dcolumn}
\usepackage{epstopdf}
\usepackage{hyperref}
\usepackage{amsmath}

\begin{document}

\title{Calculation of phase diagrams in the multithermal-multibaric ensemble}

\author{Pablo M. Piaggi}
\affiliation{Department of Chemistry and Applied Biosciences, ETH Zurich, c/o USI Campus, Via Giuseppe Buffi 13, CH-6900, Lugano, Switzerland}
\affiliation{Facolt{\`a} di Informatica, Istituto di Scienze Computazionali, and National Center for Computational Design and Discovery of Novel Materials (MARVEL), Universit{\`a} della Svizzera italiana (USI), Via Giuseppe Buffi 13, CH-6900, Lugano, Switzerland}
\author{Michele Parrinello}%
\email{parrinello@phys.chem.ethz.ch}
\affiliation{Department of Chemistry and Applied Biosciences, ETH Zurich, c/o USI Campus, Via Giuseppe Buffi 13, CH-6900, Lugano, Switzerland}
\affiliation{Facolt{\`a} di Informatica, Istituto di Scienze Computazionali, and National Center for Computational Design and Discovery of Novel Materials (MARVEL), Universit{\`a} della Svizzera italiana (USI), Via Giuseppe Buffi 13, CH-6900, Lugano, Switzerland}

\date{\today}

\begin{abstract}
From the Ising model and the Lennard-Jones fluid, to water and the iron-carbon system, phase diagrams are an indispensable tool to understand phase equilibria.
In spite of the effort of the simulation community the calculation of a large portion of a phase diagram using computer simulation is still today a significant challenge.
Here we propose a method to calculate phase diagrams involving liquid and solid phases by the reversible transformation of the liquid and the solid.
To this end we introduce an order parameter that breaks the rotational symmetry and we leverage our recently introduced method to sample the multithermal-multibaric ensemble.
In this way in a single molecular dynamics simulation we are able to compute the liquid-solid coexistence line for entire regions of the temperature and pressure phase diagram.
We apply our approach to the bcc-liquid phase diagram of sodium and the fcc-bcc-liquid phase diagram of aluminum.
\end{abstract}

\maketitle

Phase diagrams are a central tool in many areas of physics, chemistry and engineering.
They encode in a simple fashion the phase equilibria of a system.
They do so by defining regions of stability of the different phases as a function of one or more thermodynamic control variables such as the temperature, pressure, and/or composition.
In its own region of stability a phase has the minimum free energy with respect to all phases.
The determination of phase diagrams using computer simulation is crucial to understanding the properties of a given model and eventually being able to improve it.
However, calculating a phase diagram implies calculating free energy differences and this task is far from trivial.

Several methods have been devised to calculate phase diagrams using computer simulation\cite{FrenkelBook}.
The Gibbs ensemble technique developed by Panagiotopoulos\cite{Panagiotopoulos87} has proved useful to compute liquid-vapor phase diagrams as well as the properties of liquid mixtures.
Another prominent technique is thermodynamic integration\cite{FrenkelBook} that has been used in different flavors.
The variant developed by Frenkel and Ladd\cite{Frenkel84} allows calculating the free energy of solids using the Einstein crystal as reference.
Another variant of thermodynamic integration to calculate free energy differences between liquid and solids was developed by Grochola \cite{Grochola04}.
All these techniques require performing at least one Monte Carlo (MC) or molecular dynamics (MD) simulation for each point in the space of the control variables, for instance for each temperature and pressure.
Another interesting approach is that of nested sampling \cite{Partay10,Baldock16} that also allows the phase transition lines to be drawn.

In a recent work\cite{Piaggi19} we have introduced a computational approach that allows entire regions of the temperature-pressure (TP) phase diagram to be explored in a single simulation.
Applications of this method, however, were limited to one-phase regions of the TP phase diagram.
Here we propose an extension of this idea that expands significantly the scope of this type of calculation by making it possible to explore regions of the phase diagram crossed by first order phase transitions.
In this way free energy differences between phases can be calculated.
The method is based on the introduction of a bias potential that is a function of the energy, the volume, and an order parameter that describes the first order phase transition.
The bias potential is determined by using a variational principle.
We illustrate our approach with the bcc-liquid phase diagram of sodium and the fcc-bcc-liquid phase diagram of aluminum.

\section{Computing free energy differences}
In order to calculate free energy differences between the liquid and the solid $\Delta G(T,P)$ at temperature $T$ and pressure P we make use of an order parameter or collective variable (CV) $s$ which is a function of the atoms' coordinates $\mathbf{R}$.
If $s<s_0$ the configuration of the system is compatible with the liquid state and if $s>s_0$ it is compatible with the solid state.
Using $s$ the free energy difference can be expressed as
\begin{equation}
\Delta G(T,P) = -\frac{1}{\beta} \log \left ( \frac{\mathcal{P}_{T,P}(s>s_0)}{\mathcal{P}_{T,P}(s<s_0)} \right ),
\label{eq:deltaG1}
\end{equation}
where $\beta=1/k_B T$ is the inverse temperature, $k_B$ is the Boltzmann constant, and $\mathcal{P}_{T,P}(s>s_0)$ and $\mathcal{P}_{T,P}(s<s_0)$ are the probabilities of finding the system in the solid and liquid state, respectively. 
In the isothermal-isobaric ensemble at temperature $T$ and pressure $P$ the probability of finding a configuration $\mathbf{R}$ is $e^{-\beta(U(\mathbf{R})+P\mathcal{V})}/Z_{\beta,P}$ where $Z_{\beta,P}$ is the appropriate partition function and thus Eq.\ (\ref{eq:deltaG1}) can be rewritten as
\begin{widetext}
\begin{equation}
\Delta G(T,P) = -\frac{1}{\beta} \log \left ( \frac{\int\limits_{s>s_0} ds \int d\mathbf{R} \int d\mathcal{V} \: e^{-\beta(U(\mathbf{R})+P\mathcal{V})} \delta(s-s(\mathbf{R})) }{\int\limits_{s<s_0} ds \int d\mathbf{R}  \int d\mathcal{V} \: e^{-\beta(U(\mathbf{R})+P\mathcal{V})}  \delta(s-s(\mathbf{R}))}  \right ),
\label{eq:deltaG2}
\end{equation}
\end{widetext}
where $U(\mathbf{R})$ is the potential energy and $\mathcal{V}$ is the volume.
Therefore our goal is to calculate the ratio of the integrals in Eq.\ (\ref{eq:deltaG2}) in a region of the TP plane.

This ratio of integrals can be calculated using MD or MC but in order to do so the simulation must explore reversibly the liquid and the solid state.
Standard simulations are unable to do so in a reasonable time since the transformation from the liquid to the solid and vice versa is hindered by a kinetic bottleneck.
This kinetic bottleneck exists because crystallization and melting, like any first order phase transition, are triggered by nucleation.
During nucleation a cluster of the new phase emerges from the mother phase and this process has an energy cost associated with the formation of an interface.
This gives rise to a free energy barrier that must be surmounted to proceed with the transformation.
The path that we shall take here to explore reversibly the liquid and the solid states is to construct a bias potential that alters the Boltzmann probability of observing a given configuration as described below.

\begin{figure*}
\centering
\includegraphics[width=0.95\textwidth]{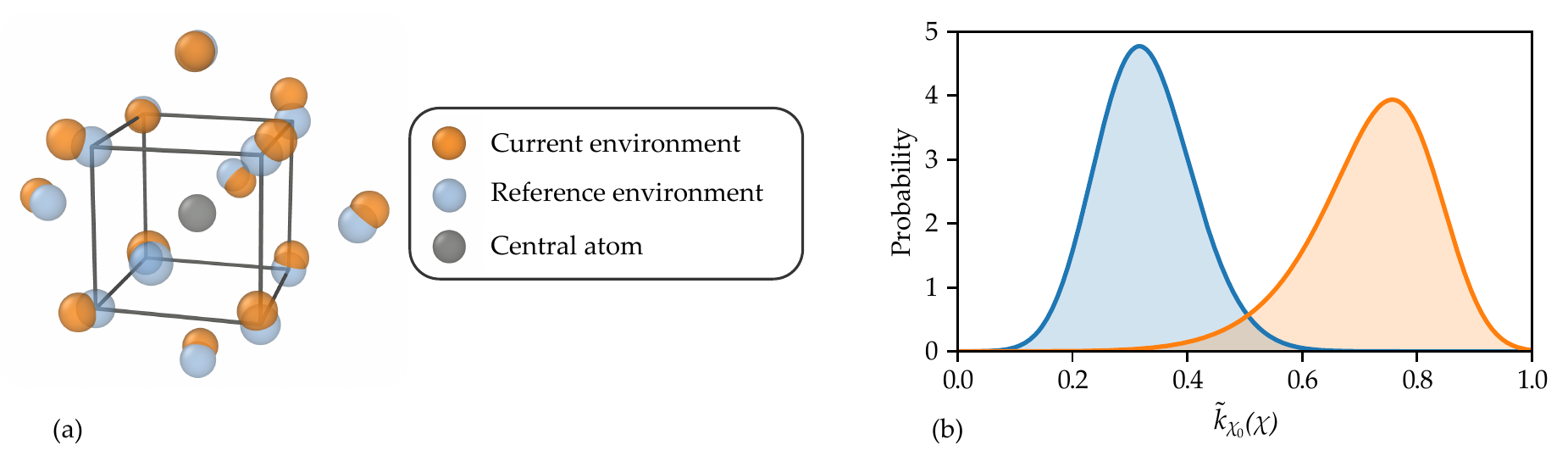}
\caption{\label{fig:Figure1} \textbf{Kernel employed to construct the order parameter.} (a) Illustration of the environments $\chi$ and $\chi_0$. The central atom is depicted in black, the reference environment or template $\chi_0$ is shown in blue, and the current environment $\chi$ is shown in orange. The current environment was extracted from a trajectory of bcc sodium at 300 K and atmospheric pressure.
                             (b) Distributions of $\tilde{k}_{\chi_0}(\chi)$ for liquid (blue) and bcc (orange) sodium at 375 K and atmospheric pressure. The distributions have a relatively small overlap.
 }
\end{figure*}

\section{Order parameter}
Another key issue is the choice of the order parameter to describe the liquid-solid transformation.
The quality of the order parameter will determine the efficiency and precision of the phase diagram calculation.
Most order parameters employed in the study of crystallization are rotationally invariant, that is to say their value is unchanged under rigid rotation.
The Steinhardt order parameters\cite{Steinhardt83,vanDuijneveldt92,Lechner08} fall in this class and so do more recent alternatives\cite{Piaggi17,Niu18}.
Here instead we generalize an approach pioneered in ref.\ \citenum{Angioletti10}  and we construct order parameters that break the rotational symmetry, thus facilitating crystallization in a preassigned direction.
If this direction is aligned to the simulation box and the appropriate number of atoms is chosen, the crystal will form without defects or artificially induced strains.
In such a way the computed free energies will be representative of the perfect crystal arrangement.

The starting point for the definition of our order parameter is the definition of a local density around an atom\cite{Bartok13,De16}.
We consider an environment $\chi$ around an atom and we define the density by
\begin{equation}
 \rho_{\chi}(\mathbf{r})=\sum\limits_{i\in\chi} \exp\left(- \frac{|\mathbf{r}_i-\mathbf{r}|^2} {2\sigma^2} \right),
 \label{eq:density}
\end{equation}
where $i$ runs over the neighbors in the environment $\chi$ and $\mathbf{r}_i$ are the coordinates of the neighbors relative to the central atom.
We now define a reference environment or template $\chi_0$ that contains $n$ reference positions $\{\mathbf{r}^0_1,...,\mathbf{r}^0_n\}$ that describe, for instance, the nearest neighbors in a given lattice.
The environments $\chi$ and $\chi_0$ are shown schematically in FIG.\ \ref{fig:Figure1}(a) taking the example of the first 14 nearest neighbors of a bcc lattice.
In the choice of $\sigma$ care must be taken that the overlap between different neighbors be negligible.
We now draw inspiration from the kernel proposed in ref.\ \citenum{Bartok13} and we compare the environments $\chi$ and $\chi_0$ using the kernel
\begin{equation}
 k_{\chi_0}(\chi)= \int d\mathbf{r} \rho_{\chi}(\mathbf{r}) \rho_{\chi_0}(\mathbf{r}) .
 \label{eq:kernel1}
\end{equation}
At variance with the kernel in ref.\ \citenum{Bartok13}, in Eq.\ (\ref{eq:kernel1}) we do not integrate with respect to all possible rotations.
If we insert Eq.\ (\ref{eq:density}) in Eq.\ (\ref{eq:kernel1}) and perform the integration analytically we obtain
\begin{equation}
 k_{\chi_0}(\chi)= \sum\limits_{i\in\chi} \sum\limits_{j\in\chi_0} \pi^{3/2} \sigma^3  \exp\left(- \frac{|\mathbf{r}_i-\mathbf{r}^0_j|^2} {4\sigma^2} \right).
 \label{eq:kernel2}
\end{equation}
The kernel is then normalized
\begin{align}
 \tilde{k}_{\chi_0}(\chi) & = \frac{k_{\chi_0}(\chi)}{k_{\chi_0}(\chi_0)} \nonumber \\
 & = \frac{1}{n} \sum\limits_{i\in\chi} \sum\limits_{j\in\chi_0} \exp\left( - \frac{|\mathbf{r}_i-\mathbf{r}^0_j|^2} {4\sigma^2} \right),
 \label{eq:kernel3}
\end{align}
such that $\tilde{k}_{\chi_0}(\chi_0) = 1$.
The kernel in Eq.\ (\ref{eq:kernel3}) is similar in spirit to the local order metric defined in ref.\ \citenum{Martelli18}.
It is instructive to calculate the distributions of $\tilde{k}_{\chi_0}$ for the solid and the liquid.
In FIG.\ \ref{fig:Figure1}(b) we show the distributions of $\tilde{k}_{\chi_0}$ for the liquid and the bcc phase of a model of sodium.

We now consider a system of $N$ atoms, and we label the environment of each atom $j$ by $\chi^j$ with $j=1,...,N$.
An order parameter $s$ for the whole system is constructed by counting the number of atoms that satisfy $\tilde{k}_{\chi_0}(\chi^j)>k_0$ where $k_0$ is chosen to be $0.5$.
This can be done in a continuous and differentiable fashion using,
\begin{equation}
s=\sum\limits_{j=1}^{N}  \frac{1-(\tilde{k}_{\chi_0}(\chi^j)/k_0)^{p}}{1-(\tilde{k}_{\chi_0}(\chi^j)/k_0)^{q}}
\end{equation}
with $p=12$ and $q=24$.
In this way $s\approx 0$ for the liquid and $s\approx N$ for the solid structure in the appropriate orientation.

The kernel can be generalized to crystal structures described as a lattice with a basis of more than one atom.
In this case there is more than one type of environment.
We consider the case of $M$ environments $X = \chi_1,\chi_2,...,\chi_M$ and we define the kernel through a best match strategy:
\begin{equation}
 \tilde{k}_X(\chi)= \frac{1}{\lambda} \log \left ( \sum\limits_{l=1}^{M}\exp \left (\lambda \: \tilde{k}_{\chi_l}(\chi) \right ) \right ).
\end{equation}
For a large enough $\lambda$ this expression will select the largest $\tilde{k}_{\chi_l}(\chi)$ with $\chi_l \in X$.
This approach can be used for instance to target the hexagonal closed packed or the diamond structure.

During an MD simulation, the crystal might form in directions not aligned with the simulation box.
In order to avoid this undesired phenomenon, we employ the following quantity,
\begin{equation}
s_{c} = \frac{Q_6-Q_6^l}{Q_6^s-Q_6^l} - \frac{\bar{k}-\bar{k}^l}{\bar{k}^s-\bar{k}^l}
\label{eq:constraint}
\end{equation}
where $Q_6$ is the global Steinhardt parameter as defined in refs.\ \citenum{vanDuijneveldt92,tenWolde96} and $\bar{k}= \sum_{i=1}^N \tilde{k}(\chi_i,\chi') / N$ is the average of the kernels defined in Eq.\ (\ref{eq:kernel3}).
The superscripts in Eq.\ (\ref{eq:constraint}) refer to the values of the order parameters in the liquid (l) and solid phases (s).
The rationale behind $s_{c}$ is as follows.
$Q_6$ is a rotational invariant parameter and if the crystal forms in any direction it increases from $Q_6^l$ to $Q_6^s$.
Instead, $\bar{k}$ is not rotationally invariant and progresses from $\bar{k}^l$ to $\bar{k}^s$ only if the crystal forms in the chosen direction.
Therefore $s_{c}$ is close to zero only if $Q_6$ and $\bar{k}$ increase simultaneously and by constraining $s_{c}$ to remain close to zero, crystals with orientations different from the desired one are avoided.
The value of $s_{c}$ can be constrained to remain close to zero by adding a restraining potential to the system's Hamiltonian.
Further details are provided in the Methods section.

\section{Multithermal-multibaric simulation}
Recently we have shown\cite{Piaggi19} that entire regions of the TP phase diagram can be sampled in a single MD simulation by constructing a bias potential $V(\mathbf{s})$ using a variational principle\cite{Valsson14}.
Within this formalism, the bias potential is determined through the minimization of the functional,
\begin{align}
\label{omega1}
\Omega [V] & =
\frac{1}{\beta} \log
\frac
{\int d\mathbf{s} \, e^{-\beta \left[ F(\mathbf{s}) + V(\mathbf{s})\right]}}
{\int d\mathbf{s} \, e^{-\beta F(\mathbf{s})}}
+
\int d\mathbf{s} \, p(\mathbf{s}) V(\mathbf{s}),
\end{align}
where $\mathbf{s}$ is a set of CVs that are a function of the atomic coordinates $\mathbf{R}$, the free energy is given within an immaterial constant by $F(\mathbf{s})=-\frac{1}{\beta}\log\int d\mathbf{R} \delta(\mathbf{s}-\mathbf{s}(\mathbf{R}) e^{-\beta U(\mathbf{R})}$,  $U(\mathbf{R})$ is the interatomic potential, and $p(\mathbf{s})$ is a preassigned target distribution.
The minimum of this convex functional is reached for:
\begin{equation}
\label{eq:optimal_bias}
V(\mathbf{s}) = -F(\mathbf{s})-{\frac {1}{\beta}} \log {p(\mathbf{s})}.
\end{equation}
which amounts to saying that in a system biased by $V(\mathbf{s})$, the distribution is $p(\mathbf{s})$.

In ref.\ \citenum{Piaggi19} in order to sample the multithermal-multibaric ensemble the potential energy $E$ and the volume $\mathcal{V}$ were chosen as CVs, and the target distribution was chosen to be
\begin{equation}
p(E,\mathcal{V})=
  \begin{cases}
    1/\Omega_{E,\mathcal{V}} & \text{if there is at least one } \beta',P' \text{ such} \\
             & \text{that } \beta' F_{\beta',P'}(E,\mathcal{V})<\epsilon \text{ with}  \\
             & \beta_1>\beta'>\beta_2 \text{ and } P_1<P'<P_2 \\
    0 & \text{otherwise}
  \end{cases}
\end{equation}
where $\Omega_{E,\mathcal{V}}$ is a normalization constant, $\beta' F_{\beta',P'}(E,\mathcal{V})$ is the free energy at temperature $T'=1/\beta' k_B$ and pressure $P'$, and $\epsilon$ is a predefined energy threshold.
In this way the final energy and volume distribution contains the energies and volumes relevant for all the desired combinations of temperatures and pressures.
Key to the success of this approach is the ability to calculate $F_{\beta',P'}(E,\mathcal{V})$ from $F_{\beta,P}(E,\mathcal{V})$ since the latter free energy can be obtained from Eq.\ (\ref{eq:optimal_bias}).
This can be done using the formula,
\begin{align}
\beta' F_{\beta',P'}(E,\mathcal{V}) = & \beta F_{\beta,P}(E,\mathcal{V}) + (\beta' - \beta) E \nonumber \\
                            & + (\beta' P' - \beta P ) \mathcal{V} + C,
\label{eq:free_energy_other_temp_press}
\end{align}
where $C$ is a constant chosen such that $\beta' F_{\beta',P'}(E_{m},\mathcal{V}_{m})=0$, and $E_m,\mathcal{V}_{m}$ is the position of the free energy minimum.

Once the bias potential has converged, the mean value of an observable in the isothermal-isobaric ensemble at temperature $T'$ and pressure $P'$ can be calculated from the multithermal-multibaric simulation using,
\begin{equation}
\langle O(\mathbf{R},\mathcal{V}) \rangle_{T',P'} = \frac{ \langle O(\mathbf{R},\mathcal{V}) w(\mathbf{R},\mathcal{V}) \rangle_{T,P,V}}
                                                             { \langle w(\mathbf{R},\mathcal{V}) \rangle_{T,P,V}},
\label{eq:reweight}
\end{equation}
where $w(\mathbf{R},\mathcal{V})=e^{(\beta-\beta')E(\mathbf{R}) + (\beta P - \beta' P') \mathcal{V}} e^{\beta V}$, $\langle \cdot \rangle_{T',P'}$ is the ensemble average in the isothermal-isobaric ensemble at temperature $T'$ and pressure $P'$, and $\langle \cdot \rangle_{T,P,V}$ is the ensemble average at temperature $T$ and pressure $P$ with bias potential $V$.

\begin{figure}
\centering
\includegraphics[width=0.9\columnwidth]{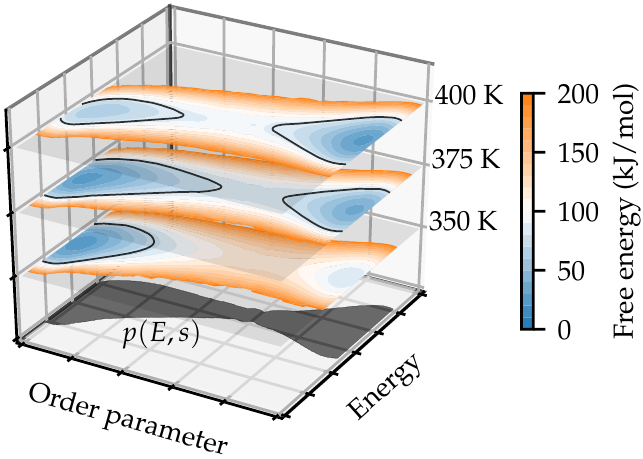}
\caption{\label{fig:Figure2} \textbf{Example of the relevant region to sample in the energy and order parameter space.}
 Three free energy surfaces at 350, 375, and 400 K are shown.
 The left and the right basins correspond to the solid and the liquid, respectively.
 The threshold of exploration $\epsilon$ is chosen to be 26 $k_B T$ and the contour line of value $\epsilon$ is shown for each of them.
 On the bottom the $p(E,s)$ is shown in gray and it represents the relevant region to sample if the temperature range 350-400 K is to be studied up to the threshold $\epsilon$.
}
\end{figure}

However, in ref.\ \citenum{Piaggi19} only regions that did not contain first order phase transitions could be studied.
During a first order phase transition the system must squeeze through a bottleneck in phase space.
Since $E$ and $\mathcal{V}$ do not describe this bottleneck properly, a bias potential constructed using these variables is incapable of driving the transformation reversibly and efficiently.
To circumvent this problem we leverage our order parameter $s$ and define a bias potential $V(E,\mathcal{V},s)$ that is a function of the energy, the volume and the order parameter.
We then define a target distribution
\begin{equation}
p(E,\mathcal{V},s)=
  \begin{cases}
    1/\Omega_{E,\mathcal{V},s} & \text{if there is at least one } \beta',P' \text{ such} \\
             & \text{that } \beta' F_{\beta',P'}(E,\mathcal{V},s)<\epsilon \text{ with}  \\
             & \beta_1>\beta'>\beta_2 \text{ and } P_1<P'<P_2 \\
    0 & \text{otherwise}
  \end{cases}
   \label{eq:target_dist}
\end{equation}
where the symbols have the same meaning as in the equations above.
The $\epsilon$ parameter must be chosen comparable to or larger than the crystallization free energy barrier such that the system is able to surmount it.
As we shall see, in the cases studied here an epsilon of 15-20 $k_B T$ seeems to be a good choice.
This will however limit the exploration of the metastable phase as discussed below for aluminum.
$\beta' F_{\beta',P'}(E,\mathcal{V},s)$ in Eq.\ \ref{eq:target_dist} can be calculated using a formula analogous to Eq.\ (\ref{eq:free_energy_other_temp_press}),
\begin{align}
\beta' F_{\beta',P'}(E,\mathcal{V},s) = & \beta F_{\beta,P}(E,\mathcal{V},s) + (\beta' - \beta) E \nonumber \\
                            & + (\beta' P' - \beta P ) \mathcal{V} + C'.
\label{eq:free_energy_other_temp_press_2}
\end{align}
where $C'$ is a constant chosen such that $\beta' F_{\beta',P'}(E_{m},\mathcal{V}_{m},s_m)=0$, and $E_m,\mathcal{V}_{m},s_m$ is the position of the free energy minimum.
The demonstration of this formula can be found in the appendix.

In order to illustrate these ideas we anticipate some of the results that will be described below.
Consider the case of the liquid-solid transformation of sodium.
We would like to study the temperature range from 350 to 400 K at atmospheric pressure and we know that the melting temperature is inside this range.
For simplicity we ignore the volume and consider a target distribution that is only a function of the energy $E$ and the order parameter $s$.
In FIG.\ \ref{fig:Figure2} we show the free energy surfaces for three temperatures 350, 375 and 400 K.
We would like to explore configurations for each temperature that are below a threshold $\epsilon=26$ $k_B T$.
The contour line of value $\epsilon$ is shown in FIG.\ \ref{fig:Figure2} for each temperature.
The points inside these contour lines have free energies lower than $\epsilon$.
In order to sample configurations relevant at temperatures from 350 K to 400 K, the union of the regions inside the contours must be sampled.
On the bottom of FIG.\ \ref{fig:Figure2} $p(E,s)$ is shown in gray and it represents the relevant region to sample if the temperature range 350-400 K is to be studied up to the threshold $\epsilon$.

Once the bias potential $V(E,\mathcal{V},s)$ that allows us to sample the distribution $p(E,\mathcal{V},s)$ has been determined, statistics can be gathered to calculate free energy differences $\Delta G(T',P')$ using Eq.\ (\ref{eq:deltaG2}) for $T_1<T'<T_2$ and $P_1<P'<P_2$.
Since a biased ensemble is being sampled, Eq.\ (\ref{eq:reweight}) must be used to calculate $\Delta G(T',P')$.
In order to use Eq.\ (\ref{eq:reweight}), we recast Eq.\ (\ref{eq:deltaG2}) as an ensemble average,
\begin{equation}
\Delta G(T',P') = -\frac{1}{\beta} \log \left ( \frac{\langle H(s-s_0) \rangle_{T',P'} }{\langle 1-H(s-s_0) \rangle_{T',P'}}  \right )
\label{eq:deltaG3}
\end{equation}
where
\begin{equation}
H(s-s_0) = \begin{cases} 1 \: \text{if} \: s>s_0 \\  0 \: \text{if} \: s<s_0 \end{cases}
\end{equation}
is the Heaviside function and $s_0$ is the value of the order parameter that defines the watershed between the liquid and the solid.
We have chosen $s_0 = N / 2$, that is to say, all configurations in which less than half the atoms are solid-like are considered liquid and those with more than half solid-like atoms are classified as solid.
This choice is not crucial since the regions of $s$ that contribute most to $\Delta G(T',P')$  are $s \approx 0$ and $s \approx N$.
Now that $\Delta G(T',P')$ is a function of ensemble averages it is easy to employ Eq.\ (\ref{eq:reweight}) to calculate it.

Using this formalism we shall study the phase diagram of sodium and aluminum.
Further details of the VES calculations are provided in the Methods section.

\begin{figure*}
\centering
\includegraphics[width=0.8\textwidth]{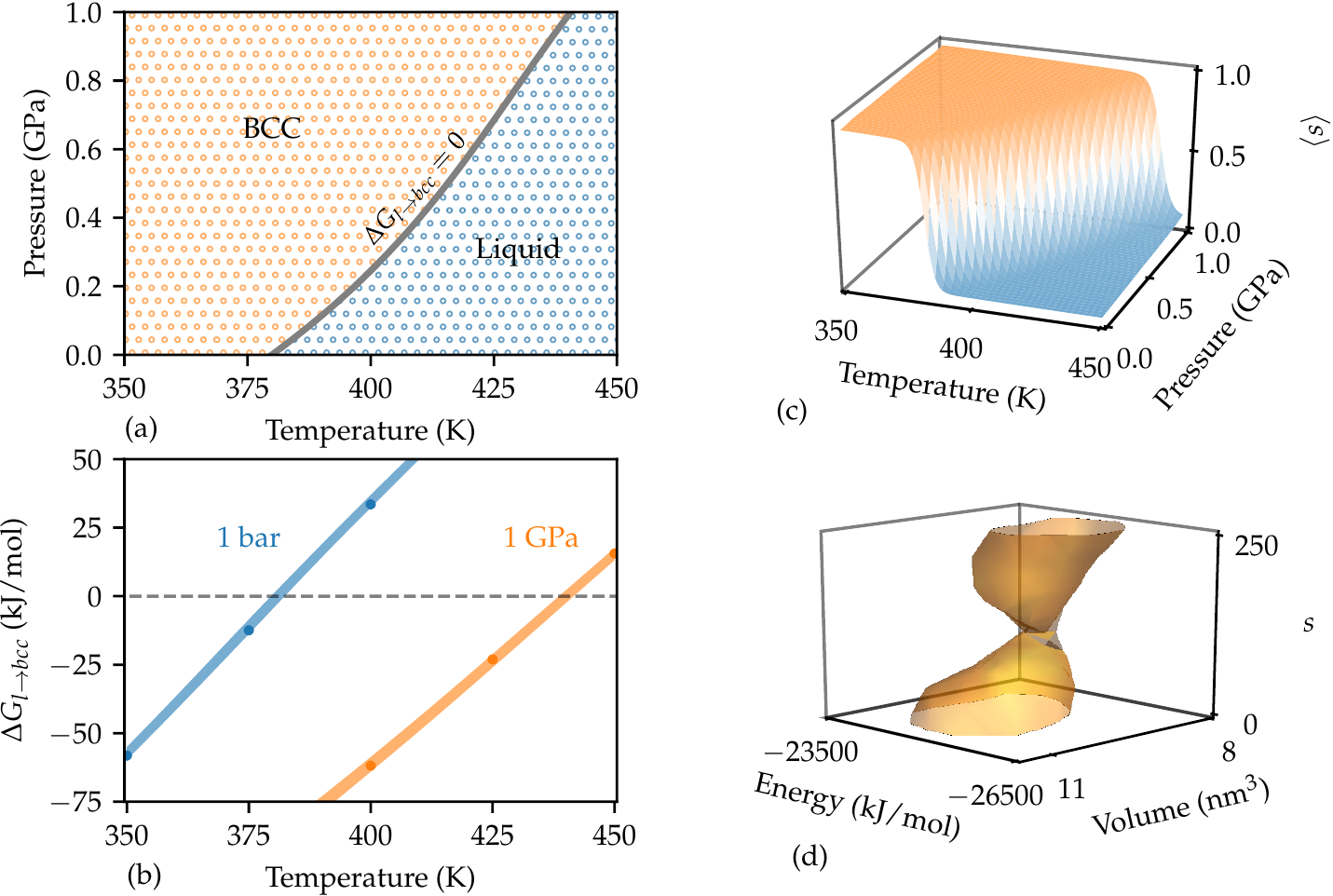}
\caption{\label{fig:Figure3} \textbf{Phase diagram of sodium.} (a) Coexistence line calculated from the multithermal-multibaric simulation.
                             (b) Free energy difference between the liquid and the bcc phase $\Delta G_{l \rightarrow bcc}$. The filled areas correspond to the estimates obtained using the multithermal-multibaric algorithm and the points are references calculated with isothermal-isobaric simulations. 
                             (c) Average of the normalized order parameter $s$ as a function of temperature and pressure.
                             (d) Target distribution $p(E,\mathcal{V},s)$ contour surface.
 }
\end{figure*}

\section{Phase diagram of sodium}
We set out to study the liquid-bcc phase diagram of a model of sodium\cite{Wilson15} using the algorithm described above.
We perform a simulation at 400 K and 0.5 GPa and we aim at gathering information of the temperature interval 350-450 K and the pressure interval 0-1 GPa.
We run the simulation for 10 ns using multiple walkers\cite{Raiteri06} and during this period the bias potential is determined such that the marginal distribution of $E,\mathcal{V},s$ is given by Eq.\ (\ref{eq:target_dist}).
After this transient the simulation is continued with fixed coefficients for a total simulation time of 100 ns.
Using this part of the simulation, we calculate the free energy difference between the liquid and the bcc phase $\Delta G_{l \rightarrow bcc}(T,P)$ using Eqs.\ (\ref{eq:reweight}) and (\ref{eq:deltaG3}).
The determination of the bias potential and the calculation of free energy differences required $2\times 10^7$ and $5\times 10^7$ force evaluations, respectively.
The results are shown as a function of temperature and pressure in FIG.\ \ref{fig:Figure3}(a).
The coexistence line $\Delta G_{l \rightarrow bcc}(T,P)=0$ is shown as a solid gray line.
Further results can be found in the supplementary information (SI).

In order to validate these results we compare some free energy isobars with references calculated from simulations in the isothermal-isobaric ensemble.
In FIG.\ \ref{fig:Figure2}(b) we show the 0 and 1 GPa isobars from the multithermal-multibaric simulation and the references.
The results are equivalent within the margins of statistical error.

It is also interesting to calculate the average of the order parameter $s$ in the isothermal-isobaric ensemble $\langle s \rangle_{T,P}$.
This is shown as a function of $T$ and $P$ in FIG.\ \ref{fig:Figure3}(c) where $\langle s \rangle_{T,P}$ has been normalized between 0 and 1.
$\langle s \rangle_{T,P}$ changes abruptly at the coexistence line giving rise to a function that resembles a cliff.
This is expected for a first order phase transition such as crystallization.
The change in  $\langle s \rangle_{T,P}$ is not perfectly discontinuous due to finite size effects.

We also found useful to plot the target distribution $p(E,\mathcal{V},s)$ once the simulation has converged.
This function represents all points of the $E,\mathcal{V},s$ space that have a free energy lower than the threshold $\epsilon$ for some $T$ and $P$ in the chosen TP interval.
In this case we have chosen $\epsilon<15$ $k_B T$.
FIG.\ \ref{fig:Figure3}(d) shows a contour surface of $p(E,\mathcal{V},s)$ and it is shaped as an hourglass.
This figure illustrates the bottleneck in the order parameter dimension where the system must squeeze through in order to transform.

\begin{figure*}
\centering
\includegraphics[width=0.8\textwidth]{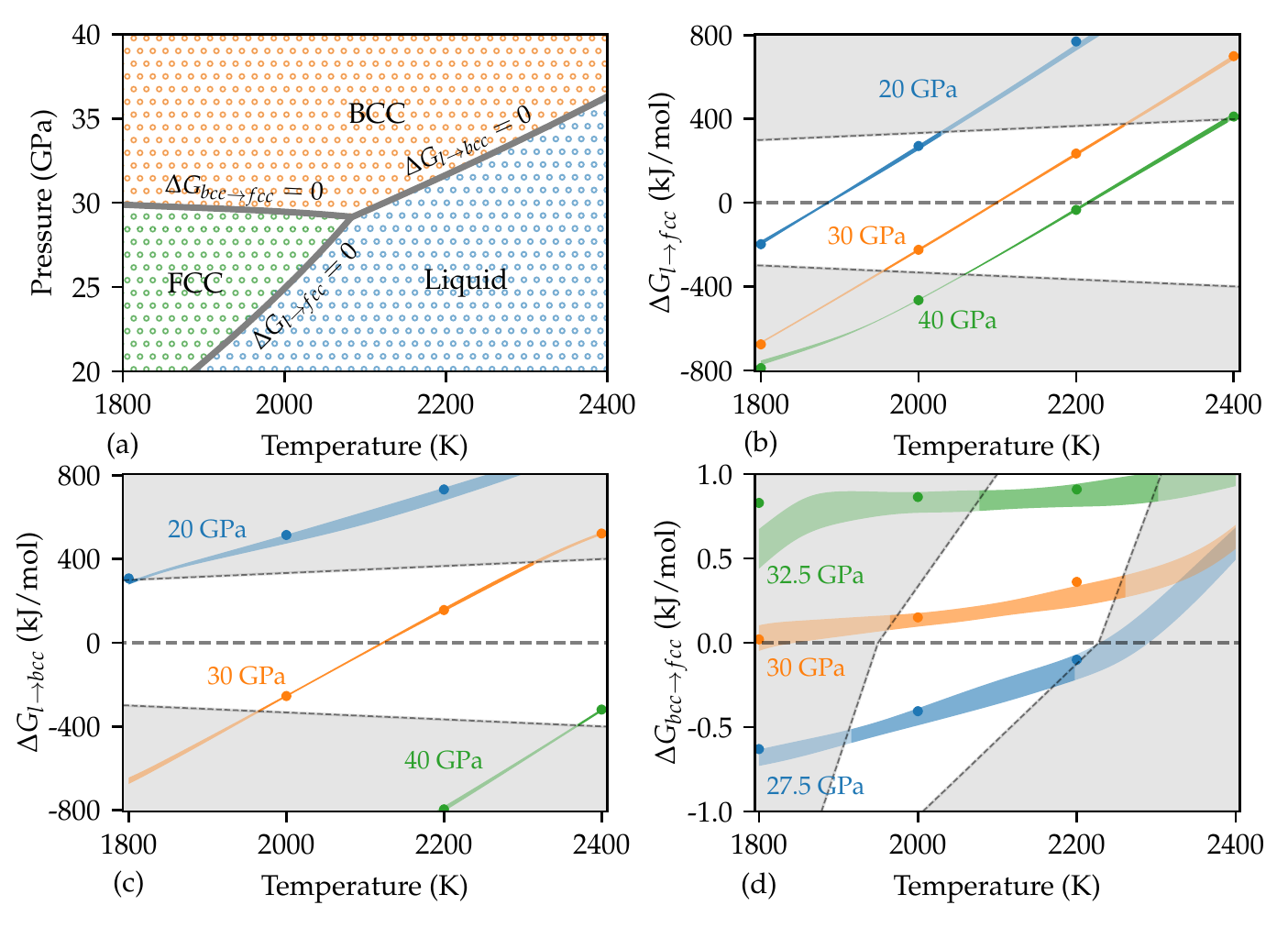}
\caption{\label{fig:Figure4} \textbf{Phase diagram of aluminum.} (a) Coexistence lines calculated from the multithermal-multibaric simulations.
                             (b) Free energy differences between the liquid and the fcc phase $\Delta G_{l \rightarrow fcc}$. 
                             (c) Free energy differences between the liquid and the bcc phase $\Delta G_{l \rightarrow bcc}$.
                             (d) Per atom free energy differences between the fcc and the bcc phase $\Delta G_{bcc \rightarrow fcc}$. References were calculated using a Bain transformation.
The filled areas correspond to the estimates obtained using the multithermal-multibaric algorithm and the filled circles are references calculated with isothermal-isobaric simulations. 
Regions shaded in gray are beyond the energy threshold chosen for the exploration.
 }
\end{figure*}

\section{Phase diagram of aluminum}
We now consider the important case of a system that exhibits polymorphism.
We choose as example a model of aluminum\cite{Mishin99} that crystallizes in the fcc phase at atmospheric pressure and in the bcc phase at high pressures.
We aim at calculating a region of the phase diagram of this model where the liquid, fcc, and bcc phase coexist.
Based on ref.\ \citenum{Baldock16} we choose to study the temperature interval 1800-2400 K and the pressure interval 20-40 GPa.
The target temperature and pressure of the thermostat and barostat were 2000 K and 30 GPa.

The approach that we shall take to obtain the phase diagram is to perform two multithermal-multibaric simulations. 
In one simulation we explore the liquid and fcc phase reversibly and we calculate the difference in free energies between the liquid and the fcc phase $\Delta G_{l \rightarrow fcc}(T,P)$ as a function of temperature and pressure.
In the other simulation we explore the liquid and the bcc phase reversibly and we calculate $\Delta G_{l \rightarrow bcc}(T,P)$.
In each simulation we optimized the bias potential for 30 ns using multiple walkers.
After this transient the bias was kept fixed and statistics were gathered for 180 ns.
Using the latter part of the simulations we calculated the coexistence lines through the formulas $\Delta G_{l \rightarrow fcc}(T,P)=0$, $\Delta G_{l \rightarrow bcc}(T,P)=0$, and $\Delta G_{bcc \rightarrow fcc}(T,P)=\Delta G_{l \rightarrow fcc}(T,P)-\Delta G_{l \rightarrow bcc}(T,P)=0$.
The determination of the bias potential and the calculation of free energy differences required $9\times 10^7$ force evaluations each.

The phase diagram obtained from these two simulations is shown in FIG.\ \ref{fig:Figure4}(a).
The triple point liquid-fcc-bcc is located at 2084 K and 29.16 GPa for the system size studied here ($\sim$250 atoms).
Further results can be found in the SI.
In order to validate our results we calculated reference free energy differences from isothermal-isobaric simulations.
In FIGs. \ref{fig:Figure4}(b) and \ref{fig:Figure4}(c) we compare the free energy differences obtained from the multithermal-multibaric simulations and the references.
The overall agreement is very good yet the accuracy seems to deteriorate for higher free energy differences.

The rationale for this behavior is as follows.
In the multithermal-multibaric simulations we employ an exploration threshold $\epsilon = 20 \: k_B T$. 
This value is chosen in order to overcome the crystallization barrier.
However, it will also affect the exploration of the metastable phase.
For instance, well inside the region of stability of the solid, the metastable liquid phase might not be explored if $|\Delta G_{l \rightarrow solid}|>\epsilon$, while if $|\Delta G_{l \rightarrow solid}|<\epsilon$ the exploration is guaranteed.
For this reason in FIGs.\ \ref{fig:Figure4}(b) and \ref{fig:Figure4}(c) we show in white the regions in which results are guaranteed and in gray the regions in which information about the metastable phase might not be available.

We also performed reference simulations in order to validate our results for the fcc-bcc free energy differences $\Delta G_{bcc \rightarrow fcc}$.
In the reference simulations we transform reversibly the bcc and the fcc phases via a Bain transformation.
This transformation was achieved by using two CVs based on the box lengths $L_x,L_y,L_z$.
The CVs are $\lambda_1=\frac{L_z/n_z-a_{fcc}}{a_{bcc}-a_{fcc}}$ and  $\lambda_2=\frac{L_x/n_x-a_{fcc}}{\sqrt{2}a_{bcc}-a_{fcc}}$ where $a_{fcc}$ and $a_{bcc}$ are the lattice constants of the bcc and fcc lattice, and $n_x$ and $n_z$ are the number of unit cells in the x and z dimensions.
The simulation is performed using $\lambda_1$ or $\lambda_2$ to construct a bias potential and under the constraints $L_x\approx L_y$ and $\lambda_1 \approx \lambda_2$.
The CV is zero in the fcc phase and one in the bcc phase, and changes smoothly between the two.
The results of these reference simulations are shown and compared with the results from the multithermal-multibaric simulations in FIG.\ \ref{fig:Figure4}(d).
The agreement is good in the region where results are guaranteed (shown in white).

\section{Conclusions}
We have presented a method for calculating liquid-solid coexistence lines in relatively large regions of the TP plane using a single MD simulation.
The method is based on a previously introduced approach to sample a multithermal-multibaric ensemble and on a novel order parameter that drives crystallization with a particular orientation in space.
We have shown the fruitfulness of our method by studying the bcc-liquid phase diagram of sodium and the fcc-bcc-liquid phase diagram of aluminum.

The results clearly show the constriction of the relevant region of the configuration space during the liquid-solid transformation.
The constriction manifests itself in the space of the energy, volume, and order parameter as an hourglass shaped region.
This is shown for sodium in FIG.\ \ref{fig:Figure3}(c), and for fcc and bcc aluminum in FIG.\ SI-3.

Another interesting result is that with a single MD simulation we obtain the ensemble average of the order parameter in a large region of the TP plane.
This function shows the discontinuity that characterizes first order phase transitions.

In the case of aluminum we performed two simulations that reproduce the fcc-liquid and the bcc-liquid transformation.
From these simulations we computed a large portion of the TP phase diagram.
We also calculated the fcc-bcc coexistence curve in the vicinity of the liquid-solid coexistence.
Our results are compatible with those obtained using nested sampling\cite{Baldock16}.

The approach presented here cannot be used directly to discover polymorphs.
We envisage that the calculation of more complicated phase diagrams could be performed in two stages.
During the first stage the possible polymorphs of a system are explored using a crystal structure prediction algorithm (see for instance refs.\ \citenum{Oganov06,Piaggi17,Piaggi18}).
Once the polymorphs are known, in a second stage one can tailor the order parameter presented here to target each structure.
The methods described in the introduction, with the exception of nested sampling, also need to know the possible polymorphs beforehand.
In our opinion, there is merit in seeking generality or specificity in this kind of method.
However, methods that target a specific structure use sampling time more efficiently in spite of not being able to predict new structures.

We emphasize that the method is not limited to liquid-solid transitions but that any first order phase transition can be studied provided that an appropriate order parameter is available.

\section{Methods}
MD simulations were performed with \textsc{LAMMPS}\cite{Plimpton95} and a development version of \textsc{PLUMED 2}\cite{Tribello14} supplemented by the VES module\cite{vescode}.
The temperature was controlled using the stochastic velocity rescaling thermostat\cite{Bussi07} and the pressure was kept constant employing the isotropic version of the Parrinello-Rahman barostat\cite{Parrinello81}.
The integration time step was 2 fs and the relaxation times of the thermostat and barostat were 0.1 and 1 ps, respectively.
Sodium and aluminum were described using the EAM potentials of Wilson et al.\ \cite{Wilson15} and Mishin et al.\ \cite{Mishin99}, respectively.
We employed systems of 250 and 256 atoms for the simulations involving bcc and fcc lattices, respectively.

The kernel for sodium was defined using $\sigma=0.065$ nm (see Eq. \ref{eq:kernel3}) and the reference environment $\chi'$ corresponded to the 14 nearest neighbors of the bcc lattice with lattice parameter $a_{bcc}=0.423$ nm.
The kernel for fcc aluminum had a $\sigma=0.04$ nm and the target environment was based on the 12 nearest neighbors of the fcc lattice with $a_{fcc}=0.38$ nm. 
The parameters of the kernel for bcc aluminum were $\sigma=0.045$ nm and $a_{bcc}=0.3$ nm.
Although the lattice constants change with temperature and pressure, the values reported above are effective in driving crystallization in the chosen regions of the TP plane.

For the multithermal-multibaric simulations a bias potential was constructed using VES and employing the energy, the volume, and $s$ as CVs.
Legendre polynomials of order 8 were used in each dimension for a total of 729 variational coefficients.
For sodium the intervals where the polynomials were defined are $-26500 <E<-23500$ kJ/mol, $8.0<\mathcal{V}<11.5$ nm, and $0<s<250$.
For bcc aluminum the polynomials were defined in the intervals  $-76000 <E<-50000$ kJ/mol, $2.8<\mathcal{V}<4.2$ nm, and $0<s<250$, and for fcc aluminum $-80000 <E<-51000$ kJ/mol, $2.9<\mathcal{V}<4.3$ nm, and $0<s<256$.
The multithermal-multibaric distribution was updated using Eqs.\ (\ref{eq:target_dist}) and (\ref{eq:free_energy_other_temp_press_2}) every 1 ns.
The exploration threshold was $\epsilon=15 \: k_B T$ for sodium and $\epsilon=20 \: k_B T$ for aluminum.
The condition in Eq.\ (\ref{eq:target_dist}) was controlled for 21 equally spaced temperatures and pressures for a total of 441 points in TP space.
The target distribution was smoothed using Gaussians with $\sigma_E=250$ kJ/mol, $\sigma_{\mathcal{V}}=0.1$ nm$^3$, and $\sigma_s=10$.
The integrals of the target distribution were performed with grid size 41x41x41.
The coefficients of the bias potential were optimized every 500 steps using the averaged stochastic gradient descent algorithm with step sizes of 10 kJ/mol for sodium and 50 kJ/mol for aluminum.
4 and 6 multiple walkers were used for sodium and aluminum, respectively.

The differences in free energy $\Delta G_{l \rightarrow solid}(T,P)$ were calculated in 161x161 equally spaced points $T_i,P_j$ in the TP plane with $i=1,...,161$ and $j=1,...,161$.
$\Delta G_{l \rightarrow solid}(T,P)=0$ was calculated by finding for each temperature $T_i$ the pressure $P_{j'}$ such that $|\Delta G_{l \rightarrow solid}(T_i,P_{j'})|<|\Delta G_{l \rightarrow solid}(T_i,P_j)|$ $\forall \: j \neq j'$ .
We thus obtained pairs $T_i,P_j'$ that define the coexistence line. 
We then fitted a B-spline to these points to create the coexistence lines shown in FIGs. \ref{fig:Figure3}(a) and \ref{fig:Figure4}(a).

The reference simulations were run at constant temperature and pressure, and the bias potential was constructed using VES employing $s$ as CV.
Legendre polynomials of order 10 were used to describe the bias potential and the well-tempered distribution with bias factor 50 was targeted\cite{Valsson15}.
The coefficients of the bias potential were optimized as described for the multithermal-multibaric case.
The target distribution was updated every 1 ns.

As described earlier, we constrained $s_{c}$ with a harmonic potential,
\begin{equation}
V(s_{c})= 
\begin{cases}
\kappa (s_c-s_c^0)^2 \quad & \mathrm{if} \quad  s_c>s_c^0 \\
0 \quad & \mathrm{otherwise}
\end{cases}
\end{equation}
with $\kappa=10^5$ kJ/mol and $s_c^0=0.1$.
In the same way as the VES potential, $V(s_{c})$ is added to the Hamiltonian and affects the dynamics through the force $- \frac{\partial V(s_c)}{\partial R_i^{\alpha} }$ where $R_i^{\alpha}$ is the $\alpha=x,y,z$ component of the position of atom $i$.
The cutoff in the calculation of $Q_6$ was 0.45 nm.

For the references calculated via the Bain transformation 256 atoms were used.
The simulation box was constrained to remain orthogonal throughout the simulation and the barostat was chosen such that $L_x$ and $L_y$ were coupled while $L_z$ was independent of the other dimensions.
A bias potential was constructed along the CV $\lambda_1$ using VES.
Legendre polynomials of order 20 defined in the interval $-0.5<\lambda_1<1.5$ were used to describe the bias potential.
The well tempered target distribution with bias factor 15 was targeted and a step size of 10 kJ/mol was used in the optimization.
Other simulation details were the same as those reported above.
Three static bias potentials were added to avoid visiting non relevant regions of configuration space.
$\lambda_1$ was constrained to the region $0<\lambda_1<1$ using the bias potentials $V_1(\lambda_1)= \kappa \lambda_1^2$ if $\lambda_1<0$ and $V_2(\lambda_1)= \kappa (\lambda_1-1)^2$ if $\lambda_1>1$ with $\kappa=10^4$ kJ/mol.
A third bias potential $V_3(\lambda_1-\lambda_2)= \kappa (\lambda_1-\lambda_2)^4$ was used to force the transformation via a smooth Bain transformation.
These biases affect the dynamics in the same way as $V(s_c)$.


\begin{acknowledgments}
We thank Robert Baldock for sharing with us his results of the coexistence lines for the model of aluminum obtained with nested sampling.
We are also grateful to Jennifer Parrinello, Michele Invernizzi and Haiyang Niu for carefully reading the manuscript.
The authors acknowledge support from the NCCR MARVEL funded by the Swiss National Science Foundation and from European Union Grant No. ERC-2014-AdG-670227/VARMET.
The computational time for this work was provided by the Swiss National Supercomputing Center (CSCS) under Project ID mr22.
Calculations were performed in CSCS cluster Piz Daint.
\end{acknowledgments}

\section*{Appendix}
We here demonstrate the formula: 
\begin{align}
\beta' F_{\beta',P'}(E,\mathcal{V},s) = & \beta F_{\beta,P}(E,\mathcal{V},s) + (\beta' - \beta) E \nonumber \\
                            & + (\beta' P' - \beta P ) \mathcal{V} + C.
\end{align}
We start from the definition of the configurational partition function in the isothermal-isobaric ensemble:
\begin{equation}
Z_{\beta,P} = \int d\mathbf{R} d\mathcal{V} e^{-\beta (U(\mathbf{R},\mathcal{V}) + P \mathcal{V})}
\end{equation}
that can be rewritten as:
\begin{equation}
Z_{\beta,P} = \int dE \: d\mathcal{V} \: ds \: e^{-\beta (E + P \mathcal{V})} N(E,\mathcal{V},s)
\end{equation}
where $N(E,\mathcal{V},s)$ is a density of states defined as,
\begin{equation}
N(E,\mathcal{V},s) = \int d\mathbf{R} \delta(E-U(\mathbf{R},\mathcal{V}) \delta(s(\mathbf{R},\mathcal{V})-s).
\end{equation}
which naturally is independent of $\beta$ and $P$.

Thus, the probability of finding a state characterized by $E$, $\mathcal{V}$ and $s$ is:
\begin{equation}
P_{\beta,P}(E,\mathcal{V},s) = N(E,\mathcal{V},s) e^{-\beta (E + P \mathcal{V})} 
\end{equation}
and using the definition of free energy $F_{\beta',P'}(E,\mathcal{V},s) = -\frac{1}{\beta} \log P_{\beta,P}(E,\mathcal{V},s)$, Eq.\ (\ref{eq:free_energy_other_temp_press_2}) is demonstrated.

\end{document}